# Corruption-free scheme of entering into contract : mathematical model

Oleg Malafeyev[1, a)] and Olga Koroleva[1, b)] and Dmitriy Prusskiy[1, c)]

and Olga Zenovich[1, d)]

[1] St. Petersburg State University, 7/9 Universitetskaya nab., St. Petersburg, 199034, Russia

**Abstract**
The main purpose of this paper is to formalise the modelling process, analysis and mathematical definition of corruption when entering into a contract between principal agent and producers. The formulation of the problem and the definition of concepts for the general case are considered. For definiteness, all calculations and formulas are given for the case of three producers, one principal agent and one intermediary. Economic analysis of corruption allowed building a mathematical model of interaction between agents. Financial resources distribution problem in a contract with a corrupted intermediary is considered. Then proposed conditions for corruption emergence and its possible consequences. Optimal non-corruption schemes of financial resources distribution in a contract are formed, when principal agent`s choice is limited first only by asymmetrical information and then also by external influences. Numerical examples suggesting optimal corruption-free agents' behaviour are presented.

**Classification Codes**: C7, D73

## 1. Introduction

Nowadays problem of corruption gets a lot of attention in both scientific and political spheres. The reason for such interest is due to harmful effects it has on society and economy. Corruption undermines the functioning of the host state and lowers the efficiency of production (Rose-Ackerman, 1999). It also hampers economic development and growth, especially since it has negative impact on investment rates (Wei and Wu, 2001). Corruption is most commonly defined as the misuse or the abuse of public office for private gain (World Bank, 1997). Recent studies are focusing more and more on specific anti-corruption programmes based on countries' experiences. A numerous research is coming from developing countries describing current economy situation and proposing anti-corruption measures.

The president of Russian Federation has set a very important task for Russia to enter into the top twenty countries in the «Ease of doing business» ranking provided by the World Bank. A high ease of doing business ranking means the regulatory environment is more conducive to the starting and operation of a local firm. Empirical research funded by the World Bank to justify their work shows that the effect of improving these regulations on economic growth is strong (World Bank, 2013).

National anti-corruption committee ran by Kirill Kabanov is conducting a research for

---
a) malafeyevoa@mail.ru
b) korolevaoa-main@ya.ru
c) dimaprusskii@mail.ru
d) olga.zenovich@gmail.com

Kremlin on corruption strategies used in Russia and other countries. As a result of this analysis, the head of the National anti-corruption committee will propose to include additional legal mechanism for fighting against corruption.

National Anti-Corruption Strategy was introduced in Russia in 2010, but there are some questions raised about its efficiency. Due to the fact that currently in Russia there is no thorough analysis of corruption practices and no thorough analysis of anti-corruption mechanisms, running anti-corruption plan does not prove to be effective.

The purpose of the conducted research, the results of which are presented in this paper, is to formalise the modelling process, analysis and mathematical definition of corruption when entering into a contract and to find optimal strategies for a corruption-free interaction between principal agent and producers.

## 2. Mathematical model of corruption-free interaction between principal agent and producers

Consider a model $\Gamma = \{I, \Phi, H\}$, where principal agent $F_0$ enters into a contract with producers $F^i$, in order to produce $q^i$ amount of goods for $s^i$ amount of money (here and below $i \in I$). During the life of the contract the prodcuers $F^i$ might face decrease (or increase) in the production levels $\beta$, depending on the amount of funds delivered by the principal agent.

The principal agent $F_0$ can contact an intermediary $\overline{F}$ in order to get information about the change in the level of production $\beta$.

In case there is an increase in production level $\beta_f$, the producer can spend less money than the amount of funds provided by the principal agent $F_0$, and thus gain additional profit. The producer $F^i$ wants to conceal this information, and hence he/she offers a bribe $z^i$ to the intermediary $\overline{F}$. In this model we consider a case of a corrupted intermediary. If there is a decrease in the production level $\beta_d$, then the provided funds are not enough to sustain production level, and hence the producer wants the official report to contain true information, and thus, he/she is required to offer bribe $z^i$ to the intermediary $\overline{F}$, as the intermediary threatens to hide this information. If the intermediary $\overline{F}$ is not aware of the above, he/she does not tend to conceal any information.

In order to stop intermediary $\overline{F}$ and the producer $F^i$ from negotiating between themselves, the principal agent $F_0$ can make a legal payment $\overline{S}$ to the intemediary $\overline{F}$. In this case the intermediary presents true information in the report $R$.

Then the principal agent makes a decision about funds distribution $S = \sum s^i$ between the producers $F^i$ based on the intermediary's $\overline{F}$ report $R$.

### 2.1 Formalisation of the corruption model $\Gamma$ when entering into a contract

Consider a model with three producers. The producers $F^1$, $F^2$ and $F^3$ together account for the following amount of goods
$$Q = q^1 + q^2 + q^3.$$

$$q^1 = \beta^1 e^1, q^2 = \beta^2 e^2, q^3 = \beta^3 e^3, \tag{1}$$

where $q^1, q^2, q^3$ - the amount of goods produced by the producers $F^1$, $F^2$ and $F^3$ respectively,

$e^1, e^2, e^3 \in R_+$ - the efforts made by the producers $F^1$, $F^2$ and $F^3$, and $\beta^i \in \{\beta_f^i, \beta_d^i\}, 0 < \beta_d^i < \beta_f^i, i = 1,2,3$ - is the change in the level of production, which is a private information held by the producers.

The payoff of the producers is defined by the formula:
$$H_{F^1} = s^1 - \phi^1(e^1) - z^1, \quad H_{F^2} = s^2 - \phi^2(e^2) - z^2, \quad H_{F^3} = s^3 - \phi^3(e^3) - z^3, \tag{2}$$

where $s^i, i = 1,2,3$ - is the size of financial funding provided by the principal agent $F_0$. The amount of funding spent by $F^i$ on the production of goods is described by the function $\varphi^i(e)$; $(\varphi^i(e))' > 0$ and $(\varphi^i(e))'' > 0$ for $i = 1,2,3$. The side payment $z^i > 0$ is offered by the producer $F^i$ to the intermediary $\bar{F}$.

The payoff function of the intermediary $\bar{F}$ is as follows:
$$H_{\bar{F}} = \bar{S} + (1-\mu)(z^1 + z^2 + z^3),$$

where $\bar{S} \geq 0$ --- is the financial funding provided by the principal agent, $\mu(z^1 + z^2 + z^3) > 0$ --- are the costs for the intermediary to conceal side payments done by the producers $F^i$.

The intermediary $\bar{F}$ examines and analyses the information about the producers $F^1$, $F^2$ and $F^3$ and either finds out about the change in the production level ($\sigma^i \in \{\beta_f^i, \beta_d^i\}, i = 1,2,3$) with probability $\pi^i$, or he/she does not find out anything ($\sigma^i = \varnothing$), i=1,2,3 with probability $(1-\pi^i)$. In case the intermediary $\bar{F}$ finds out the true situation about the production level, he/she can eaither specify this information in the report for the principal agent, or conceal it. The report sent to the principal agent $F_0$ by the intermediary $\bar{F}$ is denoted as $R = (r^1, r^2, r^3)$

$$\begin{aligned}
R \in \{&(\beta_f^1, \beta_f^2, \beta_f^3), (\beta_f^1, \beta_f^2, \beta_d^3), (\beta_f^1, \beta_d^2, \beta_f^3), (\beta_f^1, \beta_d^2, \beta_d^3), \\
&(\beta_d^1, \beta_f^2, \beta_f^3), (\beta_d^1, \beta_f^2, \beta_d^3), (\beta_d^1, \beta_d^2, \beta_f^3), (\beta_d^1, \beta_d^2, \beta_d^3), \\
&(\varnothing, \beta_f^2, \beta_f^3), (\varnothing, \beta_f^2, \beta_d^3), (\varnothing, \beta_d^2, \beta_f^3), (\varnothing, \beta_d^2, \beta_d^3), \\
&(\beta_f^1, \varnothing, \beta_f^3), (\beta_f^1, \varnothing, \beta_d^3), (\beta_d^1, \varnothing, \beta_f^3), (\beta_d^1, \varnothing, \beta_d^3), \\
&(\beta_f^1, \beta_f^2, \varnothing), (\beta_d^1, \beta_f^2, \varnothing), (\beta_f^1, \beta_d^2, \varnothing), (\beta_d^1, \beta_d^2, \varnothing), \\
&(\varnothing, \beta_f^2, \varnothing), (\varnothing, \beta_d^2, \varnothing), (\beta_f^1, \varnothing, \varnothing), (\beta_d^1, \varnothing, \varnothing), \\
&(\varnothing, \varnothing, \beta_f^3), (\varnothing, \varnothing, \beta_d^3), (\varnothing, \varnothing, \varnothing)\}
\end{aligned} \tag{3}$$

The probabilities $p^i, i = 1,2,3$ of the change in the production level being positive, together with probabilities $\pi^i$ of the intermediary $\bar{F}$ learning the information, define 32 states.

The payoff function of the principal agent is defined as follows:
$$H_{F_0} = M(Q - (T + \bar{S})), \tag{4}$$

where $Q$ - is the amount of produced goods, $T$ - are funds given to the producers $F^1$, $F^2$ and $F^3$, and $\bar{S}$ --- are funds given to the intermediary $\bar{F}$, $M$ is mathematical expectation.

The principal agent $F_0$ can not observe $\beta^i$, or $\sigma^i, i = 1,2,3$. He/she receives the report from the intermediary $R$ about the state of the companies-producers $F^1$, $F^2$ and $F^3$ and only after finds out about the amount of produced goods $q^1$, $q^2$ and $q^3$.

The principal agent $F_0$ simultaneously offers a contract to the producers $F^1, F^2$ and $F^3$ and the intermediary $\bar{F}$, in which the following is stipulated: the transmitted financial funds to the producers as a function of realised production and possibly the report, and the transmitted financial funds to the intermediary as a function of the report.

## 2.2 The construction of optimal funds distribution schemes with no corruption when entering into a contract.

### 2.2.1 The problem of funds distribution when entering into a contract under conditions of information asymmetry with no intermediary

Four states exist for each producer $F^i$:

$1: \{\beta^i = \beta^i_f, \sigma = \beta^i_f\}$, $\quad\quad 2: \{\beta^i = \beta^i_d, \sigma = \beta^i_d\}$, $\quad\quad 3: \{\beta^i = \beta^i_f, \sigma = \varnothing\}$,

$4: \{\beta^i = \beta^i_d, \sigma = \varnothing\}, i = 1, 2, 3$.

Consider the problem of funds distribution under condtion that there is no intermediary involved as a nucleus of model $\Gamma$. Define this situation in terms of the main model $\Gamma$ with $\pi^i = 0$. It follows that only the following states exist

$$13-16: \{\beta = (\beta^1, \beta^2, \beta^3), \sigma = (\sigma^1, \sigma^2, \sigma^3)\}, \tag{5}$$

where $\beta^i \in \{\beta_f, \beta_d\}, \sigma^i \in \{\varnothing\}$, for $i = 1, 2, 3$

The remaining states (1-12) have the probability of 0. The producers (but not the principal agent) observe the true change in the level of production before taking a decision on the efforts needed. By using (1), (2) and (4) the function of the principal agent can be written as

$$H_{F_0} = M \left[ \beta^1 e^1 - \phi(e^1) - H_F^1 + \beta^2 e^2 - \phi(e^2) - H_F^2 + \beta^3 e^3 - \phi(e^3) - H_F^3 \right]. \tag{6}$$

The delivery contract between the principal agent and producers is defined by the set $(e^1(\beta), e^2(\beta), e^3(\beta), H^{F^1}(\beta), H^{F^2}(\beta), H^{F^3}(\beta))$. The delivery contract should satisfy three types of conditions. The condition of rationality, $MH^i_F \geq 0, i = 1, 2, 3$ which ensures that producers stay in the project. Encouraging conditions (9) are required to make sure that the producers choose the contract designed for the real production level $\beta$. Finally, the following inequality should be satisfied $H^{F^i}_j \geq H^{F^i}_{min} < 0$ for all $i = 1, 2, 3$, which excludes the possibility of the producers going bankrupt. The principal agent solves the following problem

$$\max_{e^i_3, e^i_4, H^{F^i}_3, H^{F^i}_4} (p^1 [\beta^1_f e^1_3 - \varphi(e^1_3) - H^{F^1}_3] + (1-p^1)[\beta^1_d e^1_4 - \varphi(e^1_4) - H^{F^1}_4] + \tag{7}$$

$$+ p^2 [\beta^2_f e^2_3 - \phi(e^2_3) - H^{F^2}_3] + (1-p^2)[\beta^2_d e^2_4 - \phi(e^2_4) - H^{F^2}_4] +$$

$$+ p^3 [\beta^3_f e^3_3 - \phi(e^3_3) - H^{F^3}_3] + (1-p^3)[\beta^3_d e^3_4 - \phi(e^3_4) - H^{F^3}_4])$$

$$\begin{aligned} MH^{F^1} &= p^1 H^{F^1}_3 + (1-p^1) H^{F^1}_4 \geq 0, \\ MH^{F^2} &= p^2 H^{F^2}_3 + (1-p^2) H^{F^2}_4 \geq 0, \\ MH^{F^3} &= p^3 H^{F^3}_3 + (1-p^3) H^{F^3}_4 \geq 0 \end{aligned} \tag{8}$$

$$\phi(e^i_4) - \phi(e^i_4 \frac{\beta^i_d}{\beta^i_f}) \leq H^{F^i}_4 - H^{F^i}_3 \leq \phi(e^i_3 \frac{\beta^i_f}{\beta^i_d}) - \phi(e^i_3), i = 1, 2, 3 \tag{9}$$

$$H_3^{F^i}, H_4^{F^i} \geq H_{min}^{F^i}, i = 1,2,3 \tag{10}$$

where $H_3^{F^i} = t_3 - \varphi(e_3)$ and $H_4^{F^i} = t_4 - \phi(e_4), i = 1,2,3$ are the payoffs gained by the producers when they fulfill the contract for the real production level $\beta^i$. In case there is an increase in the production level $\beta_f^i$, the producer $F^i$ is considered to be a favourable producer; and the producer in the $\beta_d^i$ is considered to be an unfavourable type. $e_3^i$ and $e_4^i$ are the levels of the efforts, needed to produce the target amount of goods in the states 3 and 4 respectively.

Assuming that $H_j^{F^i} \geq H_{min}^{F^i} < 0$ gives the optimum, and under condition that the producers can still produce the goods even if there is a decrease in the level of production, the solution $(\bar{e}_3^i, \bar{e}_4^i, \bar{U}_3^i, \bar{U}_4^i), i = 1,2,3$ of the problem (7)-(10) is charcterised by the following formula

$$\beta_f^i = \phi'(\bar{e}_3^i) \Rightarrow \bar{e}_3^i = e_f^i, i = 1,2,3 \tag{11}$$

where $e_f^i(e_d^i), i = 1,2,3$ give the best level of efforts needed to fulfill the production plan $\beta_f^i(\beta_d^i)$. Moreover,

$$\beta_d^i = \phi'(\bar{e}_4^i) + (\Theta^i)'(\bar{e}_4^i) \frac{p^i}{1-p^i} \Rightarrow \bar{e}_4^i < e_d^i, i = 1,2,3 \tag{12}$$

where $\Theta^i(e_4^i)$ is defined as

$$\Theta^i(e_4) = H_3^{F^i} - H_4^{F^i} = \phi(e_4^i) - \phi(e_4^i \frac{\beta_d^i}{\beta_f^i}), i = 1,2,3. \tag{13}$$

The function $\Theta^i(e_4^i)$ defines a bonus, which should be offered in case production plan is achieved. $\Theta^i(e_4^i)$ is increasing with $e_4^i$, $(\Theta^i(e_4^i))' = \varphi'(e_4^i) - \varphi'(e_4^i \frac{\beta_d^i}{\beta_f^i}) \frac{\beta_d^i}{\beta_f^i} > 0$.

Based on the assumption above, $H_j^{F^i} \geq H_{min}^{F^i}$ is a necessary condition, i.e.

$$\bar{H}_4^{F^i} = H_{min}^{F^i}, i = 1,2,3,$$

hence, the expected payoff for the producers $F^1, F^2, F^3$

$$MH^{F^i} = H_{min}^{F^i} + p^i \Theta^i(\bar{e}_4^i) \geq 0. \tag{14}$$

The production plan under favourable conditions is at the optimal level with complete information, whereas it is disrupted under unfavourable conditions (it does not reach the required production level) $\bar{e}_4^i < e_d^i$.

Now define functions useful for further analysis.

Let $H_{opt1}^{F_0} = H_{opt1}^{F_01}(e_1^1, H_1^{F^1}, H_2^{F^1}) + H_{opt1}^{F_02}(e_1^2, H_1^{F^2}, H_2^{F^2}) + H_{opt1}^{F_03}(e_1^3, H_1^{F^3}, H_2^{F^3})$ be expceted payoff of the principal agent who does not know about $\beta^i$, for $\pi^i = 1, i = 1,2$. The level of efforts needed for the favourable case is fixed $e_1^i = e_f^i$, and the expected payoff of the principal agent becomes:

$$H_{opt1}^{F_0i}(e_2^i, H_1^{F^i}, H_2^{F^i}) = p^i[\beta_f^i e_f^i - \varphi(e_f^i) - H_1^{F^i}] + (1-p^i)[\beta_d^i e_2^i - \varphi(e_2^i) - H_2^{F^i}] \tag{15}$$

The first option of the best result is expressed as:

$$H_{opt1}^{F_01}(e_d^1, 0, 0) + H_{opt1}^{F_02}(e_d^2, 0, 0) + H_{opt1}^{F_03}(e_1^3, H_1^{F^3}, H_2^{F^3}) \equiv H_{opt1}^{F_0}. \tag{16}$$

Similarly, the second option of the best result is defined (for $\pi^i = 0$, $i = 1,2,3$),

$$H_{opt2}^{F_0} \equiv H_{opt2}^{F_01}(\bar{e}_4^1, H_{min}^{F^1}) + H_{opt2}^{F_02}(\bar{e}_4^2, H_{min}^{F^2}) + H_{opt2}^{F_03}(\bar{e}_4^3, H_{min}^{F^3}), \tag{17}$$

where

$$H_{opt2}^{F_0^i}(e_4^i, H_4^{F^i}) = (p^i[\beta_f^i e_f^i - \varphi(e_f^i)] + \tag{18}$$

$$+ (1-p^i)[\beta_d^i e_4^i - \varphi(e_4^i)] - [p^i \Theta^i + H_4^{F^i}]$$

### 2.2.2 Financial resources distribution problem in a contract with a corrupted intermediary

This section describes how a principal agent with asymmetrical information can increase his/her payoff by utilising information given by a more informed intermediary. Consider two scenarios: 1st scenario is a general case, in which a choice of principal agent`s resources optimal distribution is limited by only asymmetrical information; 2nd scenario is a case, in which principal agent`s choice can also be affected by limitations coming from external influence.

Assume that principal agent does not want the producer to go bankrupt, when the producer is facing a decrease in productivity. For example, even though the producer can overcome such a decrease, this can lead to the closure of factories or companies. In many countries there is a government support for the manufacturers in place. The limitations coming from external influence are represented as limitations of possible losses for the producer, i.e. he/she should not incur losses in scenarios 1 and 2. This means that one can only avoid going bankrupt when the principal agent is informed about the change in the productivity level.

### 2.2.3 Conditions for corruption emergence and its possible consequences

When $\mu < 1$ the intermediary is trying to use the information in order to get bribes. As the principal agent $F_0$ does not possess information $\sigma^i, i = 1, 2, 3$, obtained by the intermediary, the corruption can emerge in both states. In the first state the producer has a part of a revenue from the intermediary equalling the difference between the revenue of the states 3 and 1: $H_3^{F^i} - H_1^{F^i}$. The producer $F^i$ would have liked the intermediary to specify $r^i = \varnothing$ in the report in order to hide his/her profits. The intermediary $\bar{F}$ might receive a bribe for concealing this information. In state 2 the producer has a part of a revenue from the intermediary equalling the difference between the revenue of the states 2 and 4. The producer $F^i$ would have liked the intermediary $\bar{F}$ to give a fair report $r^i = \beta_d$. But the intermediary might threaten to conceal this information by specifying $r = \varnothing$ in the report. If $H_2^{F^i} - H_4^{F^i} > 0$, then there exists a risk of blackmail, i.e. the intermediary might demand the bribe for presenting true information in the report.

These 2 cases of a possible bribe have two significant differences. First difference is that the net profit for a coalition in the case of blackmail is negative, whereas the net profit in the case of offering a bribe from the producer, is positive. Second difference is that in the case of blackmail there is no influence on the report of the intermediary for the principal agent. This affects the revenue of the principal agent, because firstly, demanding the bribe reduces the potential profit of the producer and thus, it influences the individual limitation of the producer. Secondly, it creates a difference between the contractual and realised utility which can lead to non-fulfilment of the boundary condition $H^{F^i} \geq 0$.

### 2.2.4 The formation of optimal non-corruption schemes of financial resources distribution in a contract, when principal agent`s choice is limited by only asymmetrical information

Consider a problem of drawing up a contract when all parties are free to enter into this contract, and when a principal agent is only limited by asymmetrical information, i.e. when the principal agent does not observe $\beta^i$ and $\sigma^i$, $i=1,2,3$.

The first auxiliary result is as follows:

**Lemma 1.** *Let the illegal procedure of the bargain be described by a non-cooperational multi-step game: one party offers a way to share a bribe, and the other party accepts or rejects this offer, or puts forward a new offer. Then the principal agent`s decision on how to prevent producer from offering a bribe to the intermediary, does not depend on the following: 1) the direction of the bribe, i.e. to whom and who offers the bribe, 2) one of the parties makes an unambiguous offer <<take-it-or-leave-it>> or both alternatives are considered.*

The optimal mechanism of drawing up a contract is characterised by the following:

**Proposition 1.** *Corruption does not exist in the equilibrium state. The principal agent $F_0$ offers a legal payment to the intermediary $\bar{F}$ to get the report $R$, which entirely compensates for the amount of unrealised bribes. If the choice of the principal agent $F_0$ is limited by only asymmetrical information, then there is no blackmail.*

It is obvious that there is no offer of the bribe from the producer to the intermediary in the equilibrium state. This situation exists, because it is cheaper for the principal agent $F_0$ to make a legal payment $\bar{S}$ to the intermediary $\bar{F}$ (positive operational costs), rather than give up the information to the producers (or coalition). This also explains why a stimulating method of fighting against corruption always dominates. The result of the blackmail absence is less obvious: it appears that the optimal reaction to the bribe given by the producer to the intermediary excludes only a part of blackmail. In this paper we would only derive the results shown in Proposition 1.

With *Lemma 1* it is known that the principal agent`s payment $\bar{S}$, which prevents the producer from offering the bribe, should be at least equal to the amount of money which the producer could have offered to the intermediary with deduction of transaction costs. As the profits for the producer lead to the costs for the principal agent, we can write the following boundary condition in the form of equality, which prevents the producer from offering the bribe:

$$\bar{s}_1^i = (1-\mu)(H_3^{F^i} - H_1^{F^i}), \tag{19}$$

where $H_3^{F^i} = H_4^{F^i} + \Theta(e_4^i)$ due to the limitation for the producers (13). By substituting $\bar{s}_1^i$ into the principal agent`s program and by using predetermined functions $H_{opt1}^{F_0}(.), H_{opt2}^{F_0}(.)$ the following can be written

$$\max_{e_2^1, e_4^1, H_1^{F^1}, H_2^{F^1}, H_4^{F^1}, e_2^2, e_4^2, H_1^{F^2}, H_2^{F^2}, H_4^{F^2}, e_2^3, e_4^3, H_1^{F^3}, H_2^{F^3}, H_4^{F^3}} = \overset{1}{\max} + \overset{2}{\max} + \overset{3}{\max}$$

where

$$\max_{e_2^i, e_4^i, H_1^{F^i}, H_2^{F^i}, H_4^{F^i}} \pi[H_{opt1}^{F0i}(e_2^i, H_1^{F^i}, H_2^{F^i}) - (1-\mu)\left[p^1(H_4^{F^1} + \Theta(e_4^1) - H_1^{F^1}) + \right.$$

$$\left. + p^2(H_4^{F^2} + \Theta(e_4^2) - H_1^{F^2}) + p^3(H_4^{F^3} + \Theta(e_4^3) - H_1^{F^3})\right] + \quad (20)$$

$$+ (1-\pi) H_{opt2}^{F0i}(e_4^i, H_4^{F^i})$$

$$MH^{F^i} \geq 0,$$

$$\bar{s}_1^i \geq 0,$$

$$H_j^{F^i} \geq H_{min}^{F^i}, j = 1,..,4.$$

The program leads to the following first order condition: $\hat{e}_2^i$ which is determined by the equation

$$\beta_d^i = \phi'(\hat{e}_2^i) \Leftrightarrow \hat{e}_2^i = e_d^i. \quad (21)$$

In this case the optimal planned production volume is not achieved in the state 4, i.e. $\hat{e}_4^i < e_d^i$. The exact formula which determines $\hat{e}_4^i$ depends on the value of $H_{min}^{F^i}$ (see Appendix 1). Equilibrium payoffs are

$$\hat{H}_3^{F^i} = \hat{H}_4^{F^i} + \Theta(\hat{e}_4^i) \geq \hat{H}_1^{F^i} \geq H_{min}^{F^i}$$
$$\hat{H}_4^{F^i} = H_{min}^{F^i} = \hat{H}_2^{F^i} \quad (22)$$

and

$$\hat{s}_1^i = (1-\mu)[(\hat{H}_4^{F^i} + \Theta(\hat{e}_4^i)) - H_1^{F^i}] \geq 0. \quad (23)$$

The payment $\hat{s}_1^i$ - is a function of difference between producer`s payoffs in the states 3 and 1. In the point of optimality the principal agent offers a contract, in which the losses are shared between different states in a way that the difference $(H_3^{F^i} - H_1^{F^i})$ is minimal, and at the same time the expected payoffs of the producers are minimal. To achieve this it is required that, independent of the intentions of the producers, the payoff in the unfavourable situation, does not depend on the report $r^i = \beta_d^i$, i.e. $\hat{H}_2^{F^i} = H_{min}^{F^i} = \hat{H}_4^{F^i}$. Hence the optimal answer, which would prevent the producer from offering the bribe, excludes the risk of blackmail. The reason for this is the fact that the obligations of the producer are not really limited, and we would have an inequality $0 \leq \hat{H}_1^{F^i} < \hat{H}_3^{F^i} = \hat{H}_4^{F^i} + \Theta(\hat{e}_4^i)$. This means that the threat of offering the bribe by the producer, forces the principal agent to make a positive payment, even when he/she is informed. On the other hand, if the limitations are not enough to keep the producer at his level of productivity with positive profits in the state 1, then the optimal contract would lead to a negative value $H_1^{F^i}$.

The optimal contract without corruption can be described by the set of 3 elements dependent on the report $(e^i, H^{F^i}, \bar{s}^i)|_r^i$:

$$[e_f^i, H_1^{F^i}, (1-\mu)(H_{min}^{F^i} + \Theta(e_4^i)) - H_1^{F^i})]|_{r^i = \beta_f^i}$$

$$(e_d^i, H_{min}^{F^i}, 0)|_{r^i = \beta_d^i} \qquad (24)$$

$$[e_f^i, (H_{min}^{F^i} + \Theta(\hat{e}_4^i)), 0], [e_4^i, H_{min}^{F^i}, 0]|_{r^i = \varnothing}.$$

In case when information asymmetry is the only limitation when making a contract, one needs to always contact the intermediary, because the principal agent can always ignore the report when making a decision. In case when there are expenses needed to cover any side payments ($\mu > 0$), the contract made on the basis of the intermediary`s report strictly dominates the direct contract.

### 2.2.5 The formation of optimal non-corruption schemes of financial resources distribution in a contract, when principal agent`s choice is limited by asymmetrical information and external influences

Consider a scenario when in addition to information asymmetry there are also external limitations $H_1^{F^i}, H_2^{F^i}$ influencing the principal agent's choice. In this situation blackmail becomes a serious problem.

In this case, in contrary to the results obtained from Lemma1, the optimal reaction to the blackmail is sensitive to the assumptions of the bargain procedures. The reason for this is the fact that legal transfer of finances to the intermediary is not a separate process. In contrary, the legal transfer of finances including finances for producers ($H_2^{F^i} - H_4^{F^i}$) determines the joint payoff related to $\beta_d^i$, $i = 1, 2, 3$. Whether the legal transfer of money has an effect on the decision of the intermediary, depends on the way the joint payoff is shared amongst the parties.

Let assume that the intermediary makes an <<accept or reject>> offer. Then the size of the legal funds transfer, which is required to avoid the blackmail, is determined by the mechanism that forces the intermediary to conceal information in case the producer rejects the offer. The situation when the intermediary has a full bargaining power, which makes the threat of hiding information highly probable, is outside the scope of this paper.

Let assume on contrary, that the result of the illegal bargain is a joint payoff, related to $\beta_d^i, i = 1, 2, 3$, and it is shared between the parties in a way that all parties receive strictly positive sum of money. In this case, the funds transfer required to avoid the blackmail, is a function of sum of money of the intermediary.

Consider a situation where the bargain procedures can be described as an alternative offer game with a risk of failure. The <<failure>> is equivalent to the contract dependent on $r_i = \varnothing$. One of the possible interpretations is the one where the principal agent can unilaterally decide not to wait for the report, whereas intermediary and brokers are not aware of what drives the principal agent`s decision.

Analysing similar results from bargaining games, we see that the solution is the one from the Nash's problem of bargaining (how to share common cost). The sum of money, which needs to be shared is $(H_2^{F^i} - H_4^{F^i}) + \bar{s}_2^i$, where $\bar{s}_2^i$ is a legal funds transfer from the producer to the intermediary in the state 2. It is known that solution to the Nash`s problem of bargaining leads to the equal sharing of this sum. Hence, the payment $\bar{s}_2^i$ equal or more than equal to $(H_2^{F^i} - H_4^{F^i})$, confuses the intermediary and holds him/her from blackmailing the broker, because he/she can not receive a bigger payoff in the bargaining game. In reality it can be shown that, assuming infinitely small bribe (which is not considered in the paper), the principal agent can use the presence of expenses to hide side payments. If the parties have unequal bargaining powers in the game, then the minimal legal funds transfer is enough (almost surely) to avoid the demanding of the bribe, and it is equal to

$$\bar{s}_2^i = (1-\mu)[(H_2^{F^i} - H_4^{F^i})]. \tag{25}$$

The scheme, where the principal agent can prevent blackmail by means of legal funds transfer (proportional to the profits of the brokers) to the intermediary, may not be optimal, contrary to the case, when the producer is offering the bribe himself/herself.

The optimal mechanism is characterised by the following principals.

**Proposition 2**. *The limitation coming from external influence creates a risk of blackmail. At the point of optimum there is no (almost) corruption. The principal agent`s choice of using encouragements or excluding the possibility of blackmail depends on the result of illegal bargaining game.*

It is obvious, that the absence of bribe from the producer is analogous to the already considered case without limitations. Whereas, the result that there is <<almost no>> blackmail is an effect coming from the external limitation. The external limitation means that the payoff of the producer in the state 2 must not be negative, $H_2^{F^i} \geq 0$. If the producer pays the demanded bribe in the state 2, then the contract should compensate him/her for this loss, i.e. $H_{2cont}^{F^i} > 0$. Given that $\mu$ is slightly more than zero, it appears that paying compensation to the producer creates higher costs, than the legal funds transfer or exclusion of the blackmail risk. Here <<almost no>> comes from the discontinuity of the intermediary's marginal payoff at the point where there is no demanding of the bribe.

The proof of Proposition 2 was given by Lambert-Mogiliansky (1996), and full solution to the problem was also provided. Here we would provide the results for the case considered above, when the bargain's procedure is «an alternative offer with a risk of failure». The planning program is described as follows

$$\max_{e_2^i, e_4^i, H_1^{F^i}, H_2^{F^i}, H_4^{F^i}} \pi[H_{opt1}^{F_0}(e_2^i, H_1^{F^i}, H_2^{F^i}) - (1-\mu)\Big[p^1(H_4^{F^1} + \Theta(e_4^1) - H_1^{F^1}) +$$

$$+ p^2(H_4^{F^2} + \Theta(e_4^2) - H_1^{F^2}) + p^3(H_4^{F^3} + \Theta(e_4^3) - H_1^{F^3})\Big] + \tag{26}$$

$$+ (1-\pi)H_{opt2}^{F_0^i}(e_4^i, H_4^{F^i})$$

$$MH^{F^i} \geq 0$$

$$H_1^{F^i}, H_2^{F^i} \geq 0$$

$$\bar{s}_1^i, \bar{s}_2^i \geq 0$$

$$H_4^{F^i} \geq H_{min}^{F^i}, j=1,$$

where $\bar{S}_1$ and $\bar{S}_2$ are determined in (19) and (25) respectively.

First order condition for $\hat{e}_4$:

$$\beta_d^i = \phi_i'(\hat{e}_4^i) + \frac{p_i}{1-p_i}\Theta'(\hat{e}_4^i)\left(1 + \frac{\pi_i}{1-\pi_i}(1-\mu)\right), \tag{27}$$

It follows that $\hat{e}_4^i < \overline{e}_4^i$ where $\overline{e}_4^i$ is a planned volume for unfavourable type with information asymmetry (11). The reason for it is that $H_3^{F^i} = H_{min}^{F^i} + \Theta(e_4^i)$ and $\overline{s}_1^i = (1-\mu)(H_{min}^{F^i} + \Theta(e_4^i) - \hat{H}_1^{F^i})$ are increasing functions of planned volume of the producers' production when there is a failure in the state 4, $e_4^i$. The equilibrium payoffs are derived in Lambert-Mogiliansky (1996), where it is also shown that

$$\hat{H}_1^{F^i} = \hat{H}_2^{F^i} = 0$$

and

$$\hat{H}_3^{F^i} = H_{min}^{F^i} + \Theta(e_4^i).$$

It is known that producers get strictly positive payoffs, when there is asymmetry of information. But external limitation prevents the losses in case of complete information. That is why the limitation of rationality, which makes sure producers stay in the deal, is not a binding limitation at the point of optimum. That is why the sum of money $(H_3^{F^i} - H_1^{F^i})$ in case of the producer offering the bribe, can not be decreased without the increase in the payoff of the producer. Thus, we have $\hat{H}_1^{F^i} = \hat{H}_2^{F^i} = 0$. As a result, it is possible to optimally prevent the blackmail with the help of encouragements by offering a payment $\overline{s}_2^i = -(1-\mu)H_{min}^{F^i}$ where

$$(1-\mu)\pi p + (1-\pi) \leq \pi(1-p)(1-\mu). \tag{28}$$

The left part of this expression represents a limiting value of decline in the real payoff of the producer in the state 4. The first term in the expression is a marginal decline in the legal transfer of funds $\hat{s}_1^i$ and second one is a marginal decline in the producer's payoff with asymmetrical information. The right part of the expression is a marginal increase in expenses when excluding blackmail, i.e. $\hat{s}_2^i$. As function $H_0^F$ is linear over profits, the marginal net value of the producer's payoff decline in the state 4 is constant, and the principal agent here has a binary choice, i.e. either $H_4^{F^i} = H_{min}^{F^i}$ or $H_4^{F^i} = 0$. If inequality (28) is not satisfied, then the principal agent chooses to exclude blackmail by proposing a contract, in which $\hat{H}_4^{F^i} = \hat{H}_2^{F^i} = 0$. It follows that $\overline{s}_2^i = 0$.

This result can be generalised to the situations with uneven bargaining power in the bargaining game: when finding solution in the Nash bargaining problem, the minimal funds transfer required to exclude the blackmail, becomes $\hat{s}_2^i = \frac{1}{\alpha}(1-\mu)H_{min}^{F^i}$, where $\alpha$ represents the power of the producer in the bargaining game (power of the intermediary is normalised to 1). Similarly, inequality (28) becomes

$$(1-\mu)\pi p + (1-\pi) \leq \pi(1-p)(1-\mu)\frac{1}{\alpha}, \tag{29}$$

As the power of the producer $\alpha$ in the bargaining problem tends to 0, the right side of the expression becomes infinitely large. It means that the method of encouragement becomes very wasteful, that is why the principal agent uses a different method. Hence, it is not always optimal for the principal agent to prevent blackmail through encouragements, contrary to the case when the producer is offering the bribe. The most favourable method really depends on how big the sum of money the intermediary can get is. The bigger the sum the intermediary can get, the more expensive the method of encouragements is. The boundary (in terms of relative powers in the bargaining problem), on which the optimal solution switches from the method of encouragements to the method of exclusion (i.e. $\alpha$ is so that (29) is true with equality sign), is a function of model parameters $\Gamma$.

The optimal contract (on condition that (29) is satisfied) becomes:

$$[e_f^i, 0, (1-\mu)(H_{min}^{F^i} + \Theta(\hat{e}_4^i))]|_{r^i = \beta_f^i}$$

$$(e_d^i, 0, -H_{min}^{F^i})|_{r^i = \beta_d^i} \qquad (30)$$

$$[e_f^i, (H_{min}^{F^i} + \Theta(\hat{e}_4^i)), 0], [\hat{e}_4^i, H_{min}^{F^i}, 0]|_{r = \varnothing}.$$

In case when there is a corrupted intermediary in the bargain, and the contract should be satisfying external boundary conditions, then the dominance of the contract dependent on the report over a direct scheme, is not guaranteed. However, it is quite hard to deduct a sufficient condition of dominance with a general form function. Below we consider a numerical example (see Example2), in which a contract depending on the report is dominant (following some parameter specifications), if producer and intermediary have same bargaining powers in the bargaining problem. If the intermediary's bargaining power in the problem of bargaining is so high, that the principal agent prefers to exclude the threat of blackmail, then the usage of a corrupted intermediary is strictly dominated.

**Example 1.**

The following parameters are defined:

$$\beta_f^1 = 20, \beta_d^1 = 10, p^1 = 0.5, \pi^1 = 0.6, \mu = 0.4$$

$$\beta_f^2 = 21, \beta_d^2 = 7, p^2 = 0.4, \pi^2 = 0.7$$

$$\beta_f^3 = 15, \beta_d^3 = 5, p^3 = 0.5, \pi^3 = 0.8$$

Cost functions are given as:

$$\phi_1(e) = e^2$$
$$\phi_2(e) = e^2 - e$$
$$\phi_3(e) = e^2 + e$$

By using the above parameters we find that

$$\Theta_1(e) = \phi_1(e) - \phi_1(\frac{1}{2}e) = \frac{3}{4}e^2$$

$$\Theta_2(e) = 0,89e^2 - \frac{2}{3}e$$

$$\Theta_3(e) = 0,89e^2 + \frac{2}{3}e$$

$$\Theta_1'(e) = \frac{3}{2}e$$

$$\Theta_2'(e) = 1,78e - 0,67$$

$$\Theta_3'(e) = 1,78e + 0,67$$

**Solution to the funds distribution problem in the complete information game**

Solution of the game with complete information leads to the following:

$$e_f : \beta_f = \phi'(e) \Rightarrow e_f^1 = \frac{20}{2} = 10, e_f^2 = 11, e_f^3 = 7$$

$$e_d : \beta_d = \phi'(e) \Rightarrow e_d^1 = \frac{10}{2} = 5, e_d^2 = 4, e_d^3 = 2 \Rightarrow$$

$$\Rightarrow e_f^1 \beta_f^1 = 200, e_d^1 \beta_d^1 = 50, e_f^2 \beta_f^2 = 231, e_d^2 \beta_d^2 = 28, e_f^3 \beta_f^3 = 105, e_d^3 \beta_d^3 = 10$$

$$s_f^1 = \phi_1(e_f) = 100, s_f^2 = \phi_2(e_f) = 110, s_f^3 = \phi_3(e_f) = 56,$$
$$s_d^1 = \phi_1(e_d) = 25, s_d^2 = \phi_2(e_d) = 12, s_d^3 = \phi_3(e_d) = 6$$

The equilibrium payoff of the principal agent gained from first broker is:

$$H^{F_01} = M[\beta^1 e - s^1]$$

$$H^{F_01} = M[\beta^1 e - \phi(e) - H^{F_1}]$$

Similarly, from the second broker it is: $H^{F_02} = M[\beta^2 e - s^2]$

$$H^{F_02} = M[\beta^2 e - \phi(e) - H^{F_2}]$$

And from the third broker it is: $H^{F_03} = M[\beta^3 e - s^3]$

$$H^{F_03} = M[\beta^3 e - \phi(e) - H^{F_3}]$$

The optimal values are

$$H_{opt1}^{F_01} = 0,5(200-100) + 0,5(50-25) = 62,5$$

$$H_{opt1}^{F_02} = 0,4(231-110) + 0,6(28-12) = 56,8$$

$$H_{opt1}^{F_03} = 0,5(105-56) + 0,5(10-6) = 26,5$$

**Solution to the funds distribution problem in the game with information asymmetry.**

Let

$$H_{min}^{F^1} = -3$$

$$H_{min}^{F^2} = -2$$

$$H_{min}^{F^3} = -1$$

$$\beta_d^i = \phi_i'(e_4) + \frac{p^i}{1-p^i} \Theta_i'(e_4), i = 1,2,3$$

$$10 = (2+1.5)e_4^1$$

$$\bar{e}_4^1 = 2,86$$

$$\bar{e}_4^2 = 2,65$$

$$\bar{e}_4^3 = 0,88$$

As expected $\bar{e}_4^i < e_d^i$

$$H_4^{F^1} = H_{min}^{F^1} = -3$$

$$H_4^{F^2} = H_{min}^{F^2} = -2$$

$$H_4^{F^3} = H_{min}^{F^3} = -1$$

$$\hat{H}_3^{F^1} = \frac{3}{4}(2,86)^2 - 3 = 6,12 - 3$$

$$\hat{H}_3^{F^1} = 3,13$$

$$\hat{H}_3^{F^2} = 3,41$$

$$\hat{H}_3^{F3} = 0,27$$

The payoff of the principal agent under condition of asymmetrical information becomes:

$$H_{opt2}^{F_01} = 0,5(100-3,13)+0,5(28,6-8,18+3) = 60,14$$

$$H_{opt2}^{F_02} = 0,4(231-110-3,41)+0,6(7 \cdot 2,65-4,37+2) = 56,7$$

$$H_{opt2}^{F_03} = 0,5(105-56-0,27)+0,5(5 \cdot 0,88-1.65+1) = 26,2$$

$$H_{opt1}^{F_01} - H_{opt2}^{F_01} = 62,5-60,14 = 2,36$$

$$H_{opt1}^{F_02} - H_{opt2}^{F_02} = 56,8-56,7 = 0,1$$

$$H_{opt1}^{F_03} - H_{opt2}^{F_03} = 26,5-26,2 = 0,3$$

**Solution to the principal agent`s funds distribution problem with a corrupted intermediary and under the influence of external limitations.**

Consider a situation when planning of funds distribution is under the influence of external limitations.

First we find an optimal change in the production plan under unfavourable market conditions.

$$\hat{e}_4^i : \beta_d^i = \phi_i'(e_4^i) + \frac{p^i}{1-p^i}(1+\frac{\pi^i}{1-\pi^i}(1-\mu))\Theta_i'(e_4^i), i=1,2,3$$

$$10 = (2+(1+1,5 \cdot 0,6)\frac{3}{2})e_4^1$$

$$8 = 4,85e_4^2 - 1,072$$

$$5 = 6,63e_4^3 + 2,74$$

$$\hat{e}_4^1 = 2,06 < 2.86 = \overline{e}_4^1$$

$$\hat{e}_4^2 = 1,87 < 2.65 = \overline{e}_4^2$$

$$\hat{e}_4^3 = 0,34 < 0,88 = \overline{e}_4^3$$

As expected, the contract drawn up for the unfavourable market conditions in case of information asymmetry, is inevitably more skewed.

Calculate the values of the necessary legal funds transfers needed to prevent the contract from corruption.

$$\overline{s}_1^1 = 0,6(-3+\frac{3}{4}(2,06)^2)$$

$$\overline{s}_1^1 = 0,11$$

$$\overline{s}_1^2 = 0,66$$

$$\overline{s}_1^3 = 0,19$$

$$\overline{s}_2^1 = 0.6 \cdot 3 = 1,8$$

$$\overline{s}_2^2 = 1,2$$

$$\overline{s}_2^3 = 0,6$$

The principal agent`s payoff with the external limitations taken into account becomes:

$$H_{lim}^{F_01} = \pi[H_{opt1}^{F_01} - (ps_1+(1-p)s_2)] + (1-\pi)[p(\beta_f e_f - \phi(e_+^*) - (H_{min}^F + \Theta(\hat{e}_4))) + (1-p)(\beta_d \hat{e}_4 - \phi(\hat{e}_4) - H_{min}^F)]$$

$$H_{lim}^{F_01} = 0,6(62,5-0,5 \cdot 1,91) + 0,4[0,5(100-0,18) + 0,5(20,6-4,24+3)]$$

$$H_{lim}^{F_0 1} = 60,76$$
$$H_{lim}^{F_0 2} = 57,3$$
$$H_{lim}^{F_0 3} = 26,31$$

For all three producers $H_{lim}^{F_0 i} > H_{opt2}^{F_0 i}$

It is in this case when it is advised to use the intermediary, and net payoff becomes:
$$H_{lim}^{F_0 1} - H_{opt2}^{F_0 1} = 60,76 - 60,14 = 0,63$$
$$H_{lim}^{F_0 2} - H_{opt2}^{F_0 2} = 57,3 - 56,7 = 0,6$$
$$H_{lim}^{F_0 3} - H_{opt2}^{F_0 3} = 26,31 - 26,2 = 0,11$$

Based on the derived results, it is more favourable for the principal agent to make a contract with the third producer and the intermediary, and to also make legal fund transfers, which will exclude the corruption. But if the principal agent is forced to deal with all three producers (to avoid bankruptcy of either of them), then the principal agent gives the smallest possible sum of money to the producer with the lower expected payoff

**Example 2.**

Assume that the principal agent chooses to exclude the possibility of blackmail (in situation when the bargaining power of the intermediary is significantly bigger than the one of the producer). For example, the intermediary makes an offer «take-it-or-leave-it». Let us consider example based on only one producer (taking parameters from Example 1).

First we need to find an optimal change in the production plan under unfavourable market conditions $\hat{e}_4^1$, which is same as in Example 1:
$$\hat{e}_4^1 = 2,06 < 2.86 = \bar{e}_4^1$$

Now calculate the values of the legal fund transfers needed to prevent the contract from corruption.
$$\bar{s}_1^1 = 0,6(\frac{3}{4}(2,06)^2)$$
$$\bar{s}_1^1 = 1.91$$

The optimal principal agent payoff with the external limitations taken into account, becomes:
$$H_{lim}^{F_0 1} = \pi[H_{opt1}^{F_0 1} - ps_1 + (1-\pi)[p(\beta_f e_f - \phi(e_f) - \Theta(\hat{e}_4)) + (1-p)(\beta_d \hat{e}_4 - \phi(\hat{e}_4))]$$
$$H_{lim}^{F_0 1} = 0,6(62,5 - 0,5 \cdot 1,91) + 0,4[0,5(100 - 1,91) + 0,5(20,6 - 4,24)] = 59,81$$
$$H_{lim}^{F_0 1} < H_{opt2}^{F_0 1}$$

In this situation, the principal agent should not use the intermediary.

## 3. Conclusion

The president of Russian Federation has set a very important task for Russia to enter into the top twenty countries in the «Doing business» ranking provided by the World Bank by 2018. As currently analysis of corruption practices and methods of anti-corruption activities are not well documented, all government proposed solutions did not prove to be effective. The paper presents formalisation of modelling process, analysis and mathematical description of corruption problem when entering into a contract between principal agent and producers.

It is suggested to carry out further research focusing on building real models for specific projects.

### Acknowledgement
The work is partly supported by work RFBR No. 18-01-00796.